\documentstyle[sprocl,psfig]{article}
\def\Journal#1#2#3#4{{#1} {\bf #2}, #3 (#4)}
\def\APJ{\em ApJ}
\def\MN{\em MNRAS}
\def\NAT{\em Nature}
\def\be{\begin{equation}}
\def\ee{\end{equation}}
\def\bea{\begin{eqnarray}}
\def\eea{\end{eqnarray}}

\def\gtsim{\lower.5ex\hbox{$\; \buildrel > \over \sim \;$}}
\def\ltsim{\lower.5ex\hbox{$\; \buildrel < \over \sim \;$}}
\begin{document}
\title{CLUSTER EVOLUTION IN THE WIDE ANGLE ROSAT POINTED SURVEY (WARPS)}
\author{ C. A. SCHARF (GSFC/UMD), L. R. JONES (U. Birm.), E. PERLMAN (STScI), H. EBELING (IfA), G. WEGNER (Dart.), M. MALKAN (UCLA) \\ \& D. HORNER (GSFC/UMD) }
\address{Contact address: Lab. for High Energy Astrophysics, Code 662, \\
NASA/Goddard Space Flight Center, Greenbelt, MD 20771, USA}
\maketitle\abstracts{
A new catalogue of low luminosity ($L_{x} \leq 10^{44}$erg s$^{-1}$) X-ray galaxy clusters covering a redshift range of $z\sim 0.1$ to $z\sim 0.7$ has
been produced from the WARPS project.
We present the number counts of this low luminosity population
at high redshifts ($z>0.3$). The results are consistent with an
unevolving population which does not exhibit the evolution
seen in the higher luminosity cluster population. 
These observations can be qualitatively described by  self-similarly evolving dark matter and preheated IGM
models of X-ray cluster gas, with a power law index for the spectrum
of matter density fluctuations $n \geq -1$.}

\section{Galaxy cluster evolution}
The dynamical timescales of clusters of galaxies are of the order $t_{0}$, the Hubble time. Clusters are therefore still young systems. Measurements of the evolution of the X-ray luminosity function (XLF) of the most luminous ($L_{x} \geq 10^{44}$) cluste
rs in the Einstein Medium Sensitivity Survey~\cite{he} (EMSS) show evidence for some negative evolution at redshifts $z\gtsim 0.3$. Although these results allow some constraints to be put on cosmological and structure formation models they sample only the
 high end of the cluster XLF. The WARPS~\cite{sc} was
designed to extend this measurement to the faint end of the XLF, at $z>0.3$ (c.f. other similar projects~\cite{ca,ro,bu,vi}), and to further test cosmological models.

\section{The WARPS cluster sample}

Serendipitous X-ray sources were detected in ROSAT PSPC archived fields in the
0.5-2 keV band using the Voronoi Tessellation and Percolation (VTP) method~\cite{eb1,sc}. From the $16.6 $deg$^{2}$ currently surveyed ($\sim 90$ fields) a sub-sample of sources with detected flux $>3.5 \times 10^{-14}$ erg s$^{-1}$ cm$^{-2}$ (total flux 
$>5.5 \times 10^{-14}$ erg s$^{-1}$ cm$^{-2}$) was extracted, extents and corrected
fluxes were determined~\cite{sc} and complete optical followup was performed. Redshifts were obtained for most of the $ 34$ cluster candidates
which confirmed them as groups and clusters of galaxies with $10^{42}<L_{x}\leq 10^{44}$ erg s$^{-1}$ (0.5-2 keV) and $0.1\ltsim z \ltsim 0.7$. Since the detection efficencies, exposure maps and flux corrections are well understood the statistical weight
of each cluster can be accurately calculated~\cite{sc}.

\section{Testing for evolution}

\begin{figure}[t]  

\psfig{figure=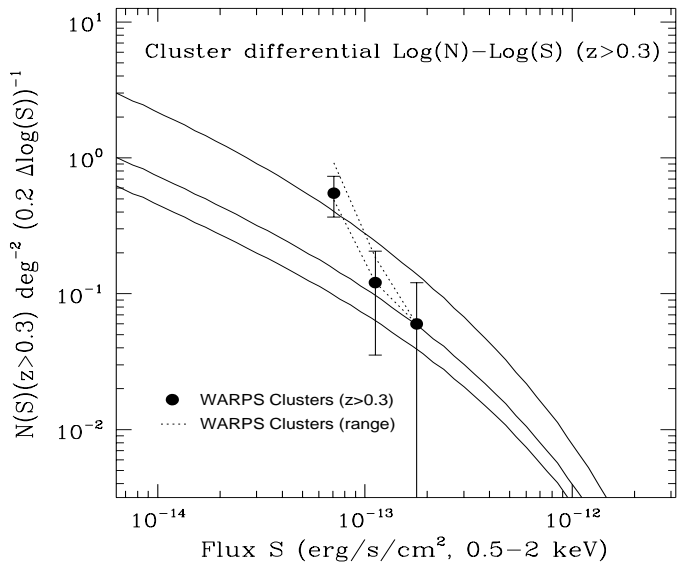,height=3.5in,width=4in} \caption{Cluster 
differential 
number counts at high redshift. Points indicate WARPS cluster counts, dotted
lines indicate possible spread in these counts due to remaining gaps in spectroscopic followup. The uppermost curve is the model prediction with no evolution ($q_{0}=0.5$), the middle curve is the prediction of negative evolution modeled as number density
 evolution, $\propto (1+z)^{-2}$, and the lowest curve is the prediction from the minimum amount of evolution seen in the EMSS (approximated as number density evolution, $\propto (1+z)^{-3}$). From Jones et al, in preparation.
 \label{fig:fig1}}
\end{figure}

In Figure~\ref{fig:fig1} the differential number counts of WARPS clusters (corrected to a uniform sky coverage) with
$z>0.3$ are presented. For comparison, at $z>0.3$ the RIXOS survey~\cite{ca} found a surface density of $0.33 \pm 0.15$ clusters deg$^{-2}$ (to a similar, slightly lower flux limit). The WARPS finds $0.84\pm 0.22$ clusters deg$^{-2}$, a factor of 2.5 time
s higher. This discrepancy is almost certainly due to different detection efficiences and data modeling.

 We have compared our results with models constructed from the low redshift cluster XLF of the ROSAT BCS~\cite{eb3}. The model predictions plotted in Figure~\ref{fig:fig1} are the expected number counts obtained by integrating the $z=0$ BCS XLF~\cite{eb3}
 to high redshift and low luminosity with $q_{0}=0.5$ under varying assumptions
about the cluster evolution.

Our conclusion is that the WARPS results show no significant evolution in the
low luminosity cluster population at $z>0.3$ (mean sample redshift $\simeq 0.4$). The amount of negative evolution allowed by the EMSS result is not seen. Interestingly, this is in qualitative agreement with the model described by Kaiser~\cite{ka}, where 
self-similar dark matter evolution combined with an initially hot IGM can reproduce the basic cluster observations. The low luminosity data here would
suggest that Kaiser's model could succeed if the power law index of the
spectrum of matter density fluctuations is $n \geq -1$. Recent X-ray observations of element abundances in cluster gas~\cite{lo} also suggest a preheated IGM.

\section*{Acknowledgments} CAS and LRJ acknowledge NRC fellowships during the major period of this work. This work was made possible by the HEASARC at NASA/Goddard Space Flight Center.


\section*{References}
\begin{thebibliography}{99} 

\bibitem{eb1}H. Ebeling \& G. Wiedenmann, Phys. Rev. E, 47, 704 (1993).
\bibitem{eb2}H. Ebeling et al, \Journal {\MN} {submitted} { }{1997}.
\bibitem{eb3}H. Ebeling et al, \Journal {\APJ} {in press} { }{1997}.
\bibitem{he}J. P. Henry et al, \Journal{\APJ} {386}{408}{1992}.
\bibitem{sc}C. A. Scharf et al, \Journal{\APJ} {March 1} { } {1997}.
\bibitem{ca}F. J. Castander et al, \Journal{\NAT} {377}{39}{1995}.
\bibitem{ro}P. Rosati et al, \Journal{\APJ} {445}{L11}{1995}.
\bibitem{bu}D. J. Burke et al, Proc. Roentgenstrahlung from the Universe, Wurzburg {1996}.
\bibitem{vi}A. Vikhlinin et al, BAAS, 188, 06.10, (1996).
\bibitem{ka}N. Kaiser, \Journal{\APJ} {383}{104}{1991}
\bibitem{lo}M. Loewenstein, R. F. Mushotzky, \Journal{\APJ} {466}{695}{1996}.
\end{thebibliography}
\end{document}